\documentstyle[epsfig]{mn}

\def\simlt{\mathrel{\rlap{\lower 3pt\hbox{$\sim$}}\raise 2.0pt\hbox{$<$}}}
\def\simgt{\mathrel{\rlap{\lower 3pt\hbox{$\sim$}} \raise 2.0pt\hbox{$>$}}}
\def\lsim{\mathrel{\rlap{\lower 3pt\hbox{$\sim$}}\raise 2.0pt\hbox{$<$}}}
\def\gsim{\mathrel{\rlap{\lower 3pt\hbox{$\sim$}} \raise 2.0pt\hbox{$>$}}}

\def\Msun{{\rm M}_{\odot}}
\def\Zsun{{\rm Z}_{\odot}}

\begin{document}
\title[Evidence for luminosity evolution of long GRBs]{Evidence for Luminosity Evolution of Long Gamma-ray Bursts in {\it Swift} Data}
\author[Salvaterra et al.]{R.~Salvaterra$^1$, C.~Guidorzi$^1$, S.~Campana$^1$,
G.~Chincarini$^{1,2}$, G.~Tagliaferri$^1$\\
$1$ INAF, Osservatorio Astronomico di Brera, via E. Bianchi 46, I-23807 Merate
(LC), Italy \\
$2$ Dipertimento di Fisica G.~Occhialini, Universit\`a degli Studi di Milano
Bicocca, Piazza della Scienza 3, I-20126 Milano, Italy }

\maketitle \vspace {7cm}

\begin{abstract}
We compute the luminosity function (LF) and the formation rate of long
gamma ray bursts (GRBs) by fitting the observed differential peak flux
distribution obtained by the BATSE experiment in two different 
scenarios: i) the GRB luminosity evolves with redshift and ii) GRBs
form preferentially in low--metallicity environments. In both cases, 
model predictions are consistent with the {\it Swift} number
counts and with the number of detections at $z> 2.5$ and $z> 3.5$.
To discriminate between the two evolutionary scenarios, we compare the
model results with the number of luminous bursts (i.e. with isotropic 
peak luminosity in excess of $10^{53}$ erg s$^{-1}$) detected by {\it Swift}
in its first three years of mission. Our sample conservatively contains 
only bursts with good redshift
determination and measured peak energy. We find that pure luminosity evolution
models can account for the number of sure identifications.
In the case of a pure density evolution scenario, models with 
$Z_{th}>0.3\;\Zsun$ are ruled out with high confidence. For lower metallicity
thresholds, the model results are still statistically consistent with 
available lower limits. However, many factors can increase the
discrepancy between model results and data, indicating that some luminosity
evolution in the GRB LF may be needed also for such low values of $Z_{th}$.
Finally, using these new constraints, we derive
robust upper limits on the bright--end of the GRB LF, showing that this cannot
be steeper than $\sim 2.6$.
\end{abstract}


\section{Introduction}

Gamma Ray Bursts (GRBs) are powerful flashes of high--energy photons occurring
at an average rate of a few per day throughout the universe. Their luminosity
is such that they can be detected up to very high redshift (the current
record is GRB~080913 at $z=6.7$; Greiner et al. 2008). The energy source of a 
GRB is believed
to be associated to the collapse of the core of a massive star in the case
of long--duration GRBs, and due to merger-- or accretion--induced collapse
for the short--hard class of GRBs (see M\'esz\'aros 2006 for a recent review).
In this paper, we limit our analysis to the class of long--duration GRBs.

The knowledge of GRBs has enormously benefited from the observations
of the {\it Swift}  satellite (Gehrels et al. 2004). Although the current 
sample of GRBs with known redshift is still too poor to allow a direct measure
of the GRB luminosity function (LF), important constraints on the cosmic 
evolution of these sources can be set on the basis of recent {\it Swift}
results.
In particular, Salvaterra \& Chincarini (2007, herethereafter SC07) showed 
that models in which GRBs are unbiased
tracer of cosmic star formation and are characterised by a constant LF
are robustly ruled out by the number of GRB detections at $z>2.5$ and $z>3.5$.
Similar conclusions were reached recently by other studies 
(e.g. Guetta, Piran \& Waxman 2005; Daigne et al. 2006; Cen \& Fang 2007; 
Kistler et al. 2008).
Moreover, they have shown that {\it Swift} data can be reproduced assuming
luminosity evolution of the GRB LF (see also Lloyd-Ronning, Fryer \& 
Ramires-Ruiz 2002; Wei \& Gao 2003; Daigne et al. 2006) and/or that GRBs form 
preferentially in 
low metallicity environments (see also Natarajan et al. 2005; 
Langer \& Norman 2006; Li 2007; Cen \& Fang 2007; Lapi et al. 2008). 

In this Letter, we derive the formation efficiency and the free parameters
describing the GRB LF by fitting the differential peak flux distribution
of BATSE GRBs in these two scenarios. We then obtain new and 
tighter constraints on the cosmic evolution of long GRBs by comparing different
models against the number of luminous (i.e. with isotropic peak luminosities
$L\gsim 10^{53}$ erg s$^{-1}$) GRBs 
detected by {\it Swift}. We also consider models with joint luminosity 
and density evolution providing a robust upper limit on the steepness of
the bright--end of the GRB LF.

This Letter is organised as follows. In Section~2, we briefly describe the
different models and the main equations used in the calculation of the LF, 
and in Section~3, we compare model results against {\it Swift} data. 
Finally, we summarise our findings in Section~4. 

\section{Model description}

The observed photon flux, $P$, in the energy band 
$E_{\rm min}<E<E_{\rm max}$, emitted by an isotropically radiating source 
at redshift $z$ is

\begin{equation}
P=\frac{(1+z)\int^{(1+z)E_{\rm max}}_{(1+z)E_{\rm min}} S(E) dE}{4\pi d_L^2(z)},
\end{equation}

\noindent
where $S(E)$ is the differential rest--frame photon luminosity of the source, 
and $d_L(z)$ is the luminosity distance. 
To describe the typical burst spectrum we adopt the
functional form proposed by Band et al. (1993), i.e. a smoothly broken power--law
with a low--energy spectral index $\alpha$, a high--energy spectral index
$\beta$, and a break energy $E_b=(\alpha-\beta)E_p/(2+\alpha)$, 
with $\alpha=-1$ and $\beta=-2.25$ (Preece et al. 2000, Kaneko et al. 2006). 
The spectrum normalisation is obtained by imposing that the 
isotropic--equivalent peak luminosity is $L=\int^{10000\,\rm{keV}}_{1\,\rm{keV}} E S(E)dE$.
In order to broadly estimate the peak energy of the spectrum, $E_p$, 
for a given $L$, we assumed
the validity of the correlation between $E_p$ and $L$ (Yonetoku et al. 2004, 
Ghirlanda et al. 2005), which is basically a diffenent expression of the 
$E_p-E_{iso}$ relation (Amati et al. 2002, Amati 2006).

\begin{equation}
E_p=337\mbox{ keV } \left(\frac{L}{2\times 10^{52}\mbox{ erg s}^{-1}}\right)^{0.49}.
\end{equation}

\noindent
Although the above
correlation has an appreciable scatter, we will show that this does not 
affect our results.

Given a normalised GRB LF, $\phi(L)$, the observed rate of 
bursts with peak flux between $P_1$ and $P_2$ is

\begin{eqnarray}
\frac{dN}{dt}(P_1<P<P_2)&=&\int_0^{\infty} dz \frac{dV(z)}{dz}
\frac{\Delta \Omega_s}{4\pi} \frac{\Psi_{\rm GRB}(z)}{1+z} \nonumber \\
& \times & \int^{L(P_2,z)}_{L(P_1,z)} dL^\prime \phi(L^\prime),
\end{eqnarray}

\noindent
where $dV(z)/dz=4\pi c d_L^2(z)/[H(z)(1+z)^2]$ is the comoving volume 
element\footnote{We adopted the 'concordance' model values for the
cosmological parameters: $h=0.7$, $\Omega_m=0.3$, and $\Omega_\Lambda=0.7$.},
and $H(z)=H_0 [\Omega_M (1+z)^3+\Omega_\Lambda+(1-\Omega_M-\Omega_\Lambda)(1+z)^2]^{1/2}$.
$\Delta \Omega_s$ is the solid angle covered on the sky by the survey,
and the factor $(1+z)^{-1}$ accounts for cosmological time dilation. 
Finally, $\Psi_{\rm GRB}(z)$ is the comoving burst formation rate. In this 
work, we model the GRB LF with a power law with an exponential
cut--off at low luminosities:

\begin{equation}\label{eq:LF}
\phi(L) \propto \left(\frac{L}{L_{\rm cut}}\right)^{-\xi} \exp \left(-\frac{L_{\rm cut}}{L}\right).
\end{equation}

SC07 have shown that models in which GRB form proportionately to the star
formation rate (SFR)
and are described by a LF constant in redshift are robustly ruled out by the 
number of GRB with sure detection at $z> 2.5$ and $z> 3.5$ during the first
two years of {\it Swift} mission. Thus, GRBs should have  
experienced some kind of evolution, being more luminous or more common in 
the past. Therefore we consider here two families of models: (i) luminosity
evolution models, where the cut--off luminosity in the GRB LF varies as 
$L_{cut}=L_0 (1+z)^\delta$ and (ii) density evolution models, where GRBs
form preferentially in galaxies with metallicity below a given threshold
$Z_{th}$. 
In the first case, the GRB
formation rate is simply proportional to the global SFR, i.e. 
$\Psi_{\rm GRB}(z)=k_{\rm GRB} \Psi_\star(z)$. We use here the recent 
determination of the SFR
obtained by Hopkins \& Beacom (2006), slightly modified to match the
observed decline of the SFR with $(1+z)^{-3.3}$ at $z\gsim 5$ suggested
by recent deep--field data (Stark et al. 2006). For the density evolution 
models,  the GRB  formation rate is obtained by convolving 
the observed SFR with the fraction of galaxies at redshift $z$ with
metallicity below $Z_{th}$ using the expression computed by Langer \& Norman 
(2006). In this scenario, the cut--off luminosity is assumed to be 
constant in redshift, i.e. $L_{cut}={\rm const}=L_0$.

Furthermore, we consider a third family of models in which both effects are 
presented: GRBs form preferentially in environments with $Z\le Z_{th}$ and are 
characterised by an evolving LF. 

\begin{table}
\begin{center}
\begin{tabular}{lcccc}
\hline
\hline
\multicolumn{5}{c}{Pure Luminosity Evolution Models} \\
\hline
$\delta$ & $k_{\rm GRB}/(10^{-8}\Msun^{-1})$ & $L_0/(10^{51}{\rm \;erg\;s}^{-1})$ & $\xi$ & $\chi_r^2$ \\
\hline
1.5 & 1.07$\pm$0.11 & 0.66$\pm$0.21 & 2.16$\pm$0.09 & 0.95 \\
2.0 & 1.01$\pm$0.10 & 0.36$\pm$0.08 & 2.08$\pm$0.06 & 0.83 \\
2.5 & 0.94$\pm$0.09 & 0.20$\pm$0.04 & 2.03$\pm$0.05 & 0.92 \\
3.0 & 0.94$\pm$0.09 & 0.10$\pm$0.02 & 1.99$\pm$0.04 & 0.84 \\
\hline
\hline
\multicolumn{5}{c}{Pure Density Evolution Models} \\
\hline
$Z_{th}/\Zsun$ & $k_{\rm GRB}/(10^{-8}\Msun^{-1})$ & $L_0/(10^{51}{\rm \;erg\;s}^{-1})$ & $\xi$ & $\chi_r^2$ \\
\hline
0.1 & 11.38$\pm$1.60 & 10.34$\pm$4.10 & 2.51$\pm$0.22 & 0.81 \\
0.2 & 4.45$\pm$0.68 & 8.40$\pm$3.58 & 2.48$\pm$0.22 & 0.78 \\
0.3 & 2.80$\pm$0.45 & 7.40$\pm$3.11 & 2.50$\pm$0.22 & 0.80 \\
\hline
\hline
\end{tabular}
\end{center}
\caption{Best--fit parameters for different models: top panel
for pure luminosity evolution and bottom panel for pure density evolution. 
Errors are at 1$\sigma$ level.}
\end{table}

\section{Luminosity \lowercase{vs} density evolution}

\begin{figure}
\begin{center}
\vskip-0.5cm
\centerline{\psfig{figure=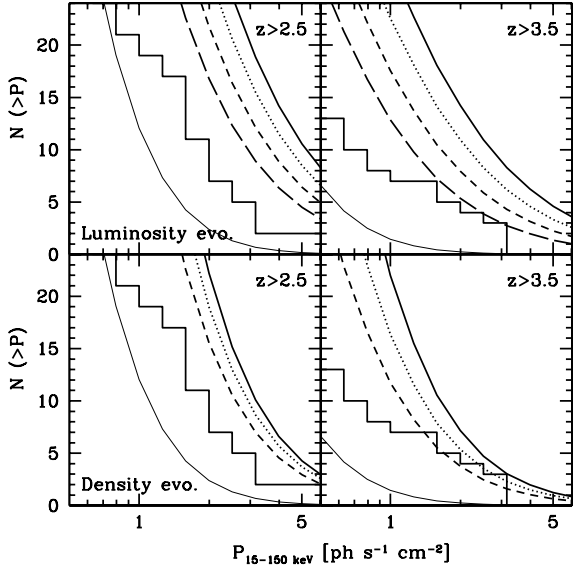,height=8cm}}
\caption{Cumulative number of GRBs at $z>2.5$ (left panels) and $z>3.5$
(right panels) as a function of the photon flux $P$. The no-evolution
case is shown with the solid thin line. Top panels refer
to luminosity evolution models: solid 
line is for $\delta=3$, dotted line for $\delta=2.5$, short--dashed line for 
$\delta=2$, and long--dashed for $\delta=1.5$. Bottom panels refer to
density evolution models: solid line is for 
$Z_{th}=0.1\;\Zsun$, dotted line is for $Z_{th}=0.2\;\Zsun$, and short--dashed 
line is for $Z_{th}=0.3\;\Zsun$. The number of bursts detected by {\it Swift} in three years is shown as solid histogram. Note that the observed detections
are lower limits, since many high-$z$ GRBs can be missed by optical follow--up
searches. A field of view of 1.4 sr for {\it Swift}/BAT is adopted.}
\vskip-0.6cm
\label{fig:z}
\end{center}
\end{figure}

The free parameters in our model are the GRB formation efficiency 
$k_{\rm GRB}$, the cut--off luminosity at $z=0$, $L_0$, and the LF power index
$\xi$. We optimised the value of these parameters by $\chi^2$
minimisation over the observed differential number counts in the 50--300 keV
band of BATSE. We use here the results by Stern et al. (2000), who
considered both triggered and non-triggered bursts and also corrected
the distribution taking into account the BATSE detector efficiency.
The best--fit parameters are reported in Table~1.
As already pointed out by SC07, it is always possible to find
a good agreement between models and data. Moreover, it is
possible to reproduce also the differential peak flux count distribution 
in the 15-150 keV {\it Swift} band using the same
GRB LF and formation efficiency obtained by fitting the BATSE data. 
Consistently with SC07, we find that the number of GRBs confirmed at $z>2.5$
and $z>3.5$
in the three years of {\it Swift} mission requires some
kind of evolution.
The results are shown in Fig. ~\ref{fig:z} for 
different evolution models. As a comparison, we show also the result for
the no-evolution model with the solid thin line. The chance probability 
associated to the no-evolution model is found to be less than $10^{-4}$ 
ensuring that this kind of model can be discarded at a very high confidence 
level and indicating the need of some kind of evolution to explain {\it
Swift} high--$z$ detections (see also SC07).

In this work we highlight a new toll to discriminate between these two 
evolution scenarios by computing the number of luminous GRBs, i.e. bursts 
with isotropic peak luminosity $L\ge 10^{53}$ erg s$^{-1}$ in the 1-10000 keV 
band. The model predictions are obtained by

\begin{equation}
\frac{dN}{dt}(>L)=\int_0^{\infty} dz \frac{dV(z)}{dz}
\frac{\Delta \Omega_s}{4\pi} \frac{\Psi_{\rm GRB}(z)}{1+z}
\int_{L_{max}(z)}^{\infty} dL^\prime \phi(L^\prime),
\end{equation}

\noindent
where $L_{max}(z)=\mbox{max}\{L,L_{0.4}(z)\}$ and $L_{0.4}(z)$ is the minimum
luminosity of a burst exploding at redshift $z$ able to trigger {\it Swift},
i.e. $P(L_{0.4},z)=0.4$ ph s$^{-1}$ cm$^{-2}$ in the 15-150 keV BAT 
band\footnote{We have assumed here a trigger threshold of 
$0.4$ ph s$^{-1}$ cm$^{-2}$ for which the {\it Swift}/BAT sample
is complete: indeed, we find that the observed differential
peak flux distribution of {\it Swift} GRBs
below this threshold is less populated than
what expected from fitting the BATSE data
(see also SC07).}.

\begin{figure}
\begin{center}
\vskip-0.5cm
\centerline{\psfig{figure=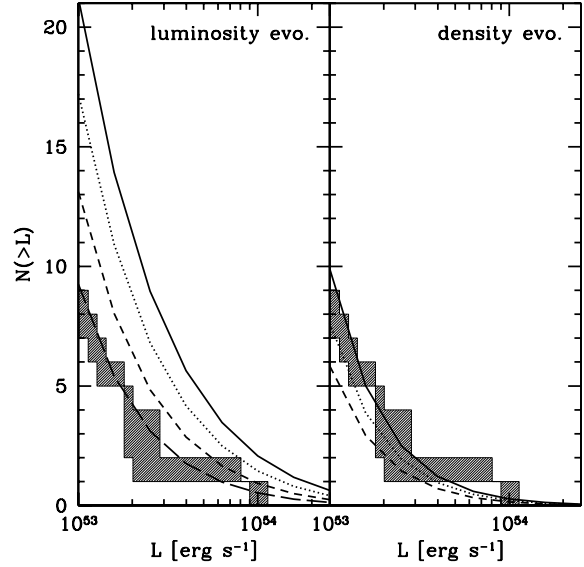,height=8cm}}
\caption{The cumulative number of luminous GRBs detected by {\it Swift} in 
three years of operations is shown with the histogram as a function of the 
isotropic equivalent peak luminosity, $L$. Shaded area takes into account  
the errors on the determination of $L$. Note that the data are to
be considered as lower limits of the real number of {\it Swift} 
detections, since only bursts with measured redshift and peak energy are 
included in our sample.
Pure luminosity (density) evolution models are shown in the left (right) 
panel. Lines as in Fig.~\ref{fig:z}.
A field of view of 1.4 sr for {\it Swift}/BAT is adopted. 
}
\vskip-0.6cm
\label{fig:bright}
\end{center}
\end{figure}

We compare model predictions with the number of bright bursts detected
by {\it Swift} in three years of mission. 
Conservatively, our data sample contains only bursts with a 
good redshift measurement 
and whose peak energy was measured or well constrained by 
{\it Swift} itself or other satellites (such as HETE-2 or Konus-Wind).
We find nine GRBs\footnote{Six out of the nine GRBs included in our sample
are reported in Rossi et al. (2008): GRB~050401, GRB~050603, GRB~060927, 
GRB~061007, GRB~061121, and GRB~071020. In this work we use the peak luminosity
computed on a 1 s timescale. We add to these other three bursts
that are not present in Rossi et al. (2008) compilation: 
GRB~050505 with $L=1.6\pm1.0\times 10^{53}$ erg s$^{-1}$, 
GRB~060210  with $L=4.6\pm2.8\times 10^{53}$ erg s$^{-1}$, and
GRB~060124 with $L=1.1\pm0.1\times 10^{53}$ erg s$^{-1}$ (Romano et al.
2006).  We note that the peak fluxes
of all of the nine bursts considered in our sample
are well above the assumed trigger threshold of {\it Swift}/BAT.}
detected by {\it Swift} with $L\ge 10^{53}$ erg s$^{-1}$ in three years
of mission.
We want to stress here that this number represents a conservative lower 
limit on the real number of bright GRBs detected, since some luminous bursts 
without $z$ and/or $E_p$ can be present in the {\it Swift} catalogue.
In particular, we note that a good redshift determination is obtained for 
$\sim 1/3$ of {\it Swift} bursts and for only a fraction of these we have a
well constrained $E_{p}$. The cumulative distribution of the known bright
bursts is shown in Fig.~\ref{fig:bright}. The shaded
area takes into account errors on the determination of $L$. 
The model results
for the pure luminosity (density) evolution models are plotted in the 
left (right) panel of Fig.~\ref{fig:bright}. We have also checked that
our findings do not depend on the assumed $L-E_p$ correlation:
considering a mean break energy (as done in SC07) for all bursts
does not change significantly our results.

For the pure luminosity evolution scenario, all models here 
considered predict a large number of bright bursts to be detected by {\it
Swift}. Indeed, they can easily account for the observed number of 
bursts with $L>10^{53}$ erg s$^{-1}$. 
On the other hand, models in which GRB formation
is confined in low--metallicity environments seems to fall short to account 
for the observed bright GRBs for $Z_{th}>0.1$. 
In particular, for $Z_{th}=0.3\;\Zsun$, as required by collapsar models 
(MacFadyen \& Woosley 1999; Izzard, Ramirez-Ruiz \& Tout 2004), only $\sim 6$ 
bursts with $L\ge 10^{53}$ erg s$^{-1}$ should have been detected in three 
years of observations.

\begin{figure}
\begin{center}
\vskip-0.5cm
\centerline{\psfig{figure=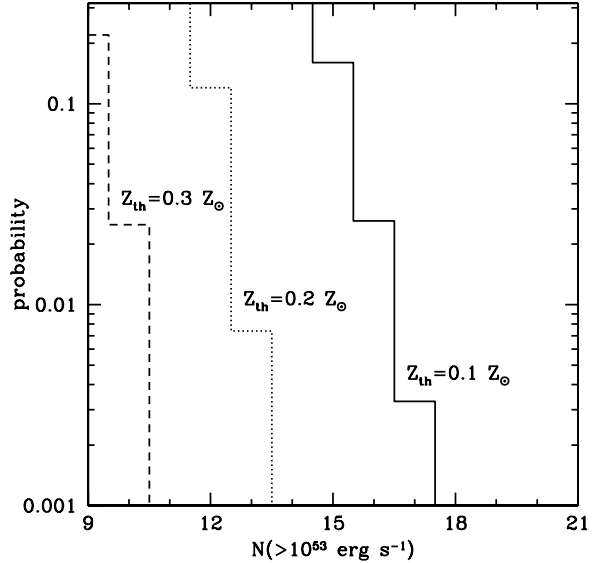,height=8cm}}
\caption{Chance probability to reproduce the
differential peak flux  distribution of BATSE GRBs given a model that
predicts $N$ bursts with  $L\ge 10^{53}$ erg s$^{-1}$ in 3 years of {\it
Swift} operations.  Different pure density evolution models are
plotted with different lines: dashed line for $Z_{th}=0.3\;\Zsun$,
dotted line for $Z_{th}=0.2\;\Zsun$, and solid line for $Z_{th}=0.1\;\Zsun$.
The probability rapidly decreases with the number of bright bursts so 
that, even a few more bright bursts to be added to our conservative sample 
would require some degree of luminosity evolution.
}
\vskip-0.6cm
\label{fig:prob}
\end{center}
\end{figure}

In order to test the statistical significance of our findings, 
we explore the parameter space around the  best fit
parameters looking for triple ($k_{\rm GRB}$, $\xi$, $L_0$) compatible with
the sure bright burst identifications. Among these, we choose the triple
that give the best agreement with the differential peak flux distribution
of BATSE GRBs. Then, the null hypothesis test gives us the confidence level 
at which we can discard the considered model. For $Z_{th}=0.3\;\Zsun$, we
find that null hypothesis is satisfied being, in the best case, the chance 
probability of $\sim0.22$. However, since we are dealing with strong lower
limit on the real number of bright burst detections, we have also to consider
the case in which some luminous bursts are hidden among the GRBs missing
redshift and/or $E_p$ measurement. We find that the null hypothesis 
probability decreases rapidly with the number of bright burst detections and
already for 10 bright bursts it drops down to the percent level (see 
Fig.~\ref{fig:prob}). So, although 
the $Z_{th}=0.3\;\Zsun$ scenario cannot formally be discarded by available data, some
degree of luminosity evolution in the GRB LF is suggested if just a few more
bright bursts
had to be added to our measured sample. For $Z_{th}=0.1$ (0.2) $\Zsun$,
the models can account up to 16 (12) bright bursts in the whole {\it Swift}
data sample, i.e. just 7 (3) 'hidden' bright bursts (Fig.~\ref{fig:prob}). For the pure evolution 
scenario, we find that even a relatively 
large population of bright bursts hidden in the {\it Swift} catalogue of 
bursts without good redshift and/or $E_p$ measurement, can still be accounted
for.

\begin{figure}
\begin{center}
\vskip-0.5cm
\centerline{\psfig{figure=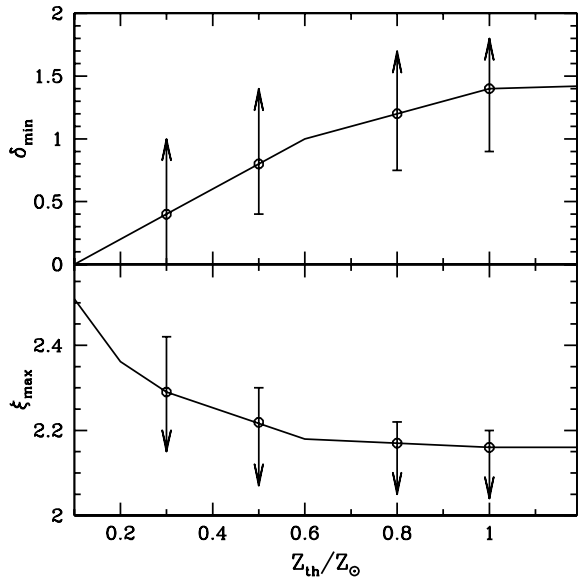,height=8cm}}
\caption{{\it Top panel}: lower limit on the luminosity evolution 
parameter, $\delta$, for different threshold metallicity $Z_{th}$. 
Error bars correspond to a null hypothesis probability, that the model 
reproduces the differential peak flux distribution of BATSE 
GRBs, equal to 1\%. {\it Bottom panel}: corresponding upper limit 
on the bright--end power index of the GRB LF, $\xi$.}
\vskip-0.6cm
\label{fig:limits}
\end{center}
\end{figure}

To complete our analysis, we then consider models in which both luminosity and density evolution are
present. For any value of the threshold metallicity $Z_{th}$, we derive the
minimum luminosity evolution required to match the number of known GRBs with 
$L\ge 10^{53}$ erg s$^{-1}$ in the 3--year catalogue (top panel of 
Fig.~\ref{fig:limits}). The error bars correspond to a null hypothesis 
probability of 1\% that the model reproduces the differential peak flux 
distribution of BATSE GRBs. We find that for $Z_{th}>0.3\;\Zsun$, the
available constraints require luminosity evolution in the GRB LF. 
Below this threshold, models charaterised by a constant LF can account 
for the number of known luminous bursts considering the error bars. As
already pointed out, the existence of a few bright bursts hidden among GBRs
without a sure redshift and/or $E_p$ measure (more than 2/3 of the whole
{\it Swift} catalogue), would imply the need of luminosity evolution even
for lower value of $Z_{th}$.
 Moreover, note here that the existence of a distinct 
population of long low-luminosity GRBs at low-z (see e.g. Guetta 
\& Della Valle 2007) strengthens our conclusions. In the form of the adopted 
LF (see eq.~(\ref{eq:LF})), we do not include this population in our
calculations. Should this population be statistically significant and present
at all redshifts, the faint end of the LF would be more populated and
all model predictions would be shifted 
towards lower values, increasing the discrepancy between model results and 
data. Since for pure density evolution models the LF cut-off luminosity is 
considerably larger than for pure luminosity evolution models, the former
models would be more severely affected by the existence of a large 
population of underluminous bursts at any redshift.
In conclusion, although we cannot rule out at a high confidence level pure
density evolution models with $Z_{th}<0.3\;\Zsun$, available data  
suggest some luminosity evolution in the GRB LF with redshift.

Finally, we constrain the steepness of the bright--end of the GRB LF (bottom
panel of Fig.~\ref{fig:limits}). 
Very steep LF are robustly 
ruled out in every model considered. We find that the maximum value of the
index $\xi$ have is $\lsim 2.6$. For the pure luminosity 
evolution models, we find $\xi<2.2$.
 As we already pointed out, these limits could be further constrained  
to explain current {\it Swift} data due to the possible existence of
bright bursts missed by our conservative selection and of a relatively
large population of faint bursts not considered here.

\section{Conclusion}

We have computed the luminosity function  and the formation rate of long
GRBs by fitting the observed differential peak flux
distribution obtained by the BATSE satellite in two different 
scenarios: i) the GRB luminosity evolves with redshift and ii) GRBs
form preferentially in low--metallicity environments. In both cases, 
model predictions are consistent with the {\it Swift} number
counts and with the number of detections at $z> 2.5$ and $z> 3.5$.
To discriminate between the two evolutionary scenarios, we compared
the model results against the number of luminous bursts, i.e. with peak luminosity
in excess of $10^{53}$ erg s$^{-1}$, detected by {\it Swift}. 
Conservatively, our data sample contains only bursts with good redshift
determination and measured peak energy. We find that models in which GRBs
are characterised by a constant LF (i.e. for pure density evolution models) 
are disfavoured as they underpredict the number of luminous GRBs.
{\it Swift} data can be explained assuming that the GRB luminosity
evolves with redshift. Although we cannot discard pure density evolution
models with $Z_{th}<0.3\;\Zsun$ on the basis of the current sample, the 
existence
of a few bright GRBs missed by our conservative selection criteria and/or
of a relatively large population of faint GRBs would require some luminosity
evolution in the GRB LF even for such low values of $Z_{th}$. 
On the other hand, pure luminosity evolution scenarios can account more easily
for a large number of burst detections with $L>10^{53}$ erg s$^{-1}$.
In this work, we derive lower limits to the luminosity evolution of
the GRB LF with redshift for different values of the metallicity threshold
for the GRB formation. Moreover, we use these constraints to set 
a robust upper limit on the bright--end of GRB LF. We find that the number of 
bright GRBs detected by {\it Swift} implies that this cannot be very 
steep: $\xi\lsim 2.2$ (pure luminosity evolution) and $\xi\lsim 2.6$ (for
$Z_{th}<0.3\;\Zsun$).

In conclusion, we find that available {\it Swift} observations  
point towards a scenario where GRBs were more luminous in the past. 
Although current data sample of bright GRBs with good redshift and 
$E_{p}$ determination is still very poor, our findings show that 
these data can be used to set important constraints on the cosmic evolution 
of GRBs and on the steepness of their LF. 

Finally, the new constraints on the GRB LF allow us to derive robust
lower limits on the number of bursts detectable by {\it Swift} at very 
high redshift. Assuming a trigger threshold $P_{lim}=0.4$ ph s$^{-1}$ 
cm$^{-2}$, at least $\sim 5-10$\% of
all detected GRBs should lie at $z\ge 5$, where the lower (upper) value 
refers to a pure luminosity evolution (pure density evolution with 
$Z_{th}=0.1\;\Zsun$) model. 
Among these, $>1-3$ GRB yr$^{-1}$ should be detected at $z\ge 6$. These lower
bounds double by lowering the {\it Swift} trigger threshold by a 
factor of two (Salvaterra et al. 2008). 
These results are consistent with the lower limits on the number of high--$z$
detections obtained by Salvaterra et al. (2007, 2008) using redshift
distribution constraints.

\end{document}